\documentclass{article}
\pdfoutput=1

\usepackage{arxiv}
\usepackage{amsmath}
\DeclareMathOperator*{\argmax}{arg\,max}

\usepackage{algorithm}
\usepackage{algpseudocode}

\usepackage{adjustbox}
\usepackage{amsfonts}
\usepackage{amssymb}
\usepackage{amsbsy}
\usepackage[T1]{fontenc}
\usepackage{textcomp}
\usepackage{graphicx}
\usepackage{lineno}
\usepackage[normalem]{ulem}
\usepackage{color}
\definecolor{ruby}{rgb}{0.6,0,0.3}
\definecolor{darkruby}{rgb}{0.4,0,0.2}
\definecolor{whitegold}{rgb}{0.8667,0.8510,0.7647}
\definecolor{lightgray}{rgb}{0.7,0.7,0.7}
\usepackage{xcolor,colortbl}
\usepackage{soul}

\usepackage{epstopdf}
\usepackage{multirow}
\usepackage{xparse}
\ExplSyntaxOn
\NewDocumentCommand{\smallcaps}{m}
 {
  \tl_set:Nn \l_tmpa_tl { #1 }
  \regex_replace_all:nnN
   { ([0-9]+) } 
   { \c{resizedigit}\cB\{ \1 \cE\} } 
   \l_tmpa_tl
  \textsc{ \tl_use:N \l_tmpa_tl }
 }

\ExplSyntaxOff

\usepackage[hidelinks]{hyperref}
\usepackage{url}
\usepackage{xr} 

\graphicspath{{./}}
\newcommand{\ignore}[1]{}
\newcommand{\revisit}[1]{}
\renewcommand{\vec}[1]{\boldsymbol{#1}}
\newcommand{\mat}[1]{\mathbf{#1}}
\newcommand{\sep}[0]{ $\cdot$ }
\newcommand\defeq{\mathrel{\overset{\makebox[0pt]{\mbox{\normalfont\tiny def}}}{=}}}

\usepackage{tikz}
\usepackage{tikzpagenodes}
\usepackage{background}
\backgroundsetup%
{   angle=0,
    opacity=1,
    scale=1,
    contents=%
    {   \begin{tikzpicture}[remember picture,overlay]
            \node[text=gray,rotate=90,above=10mm] at (current page.east) {\scriptsize\color{lightgray}\fbox{\begin{minipage}{195mm}{\scriptsize\textcopyright 2026 IEEE.  Personal use of this material is permitted.  Permission from IEEE must be obtained for all other uses, in any current or future media, including reprinting/republishing this material for advertising or promotional purposes, creating new collective works, for resale or redistribution to servers or lists, or reuse of any copyrighted component of this work in other works.}\end{minipage}}}; 
        \end{tikzpicture}
    }
}

\title{Effective information gathering for ore estimation,\\evaluation and perspectives on adaptive sampling}

\date{\vspace{-5mm}\begin{tabular}{ll}
\textbf{To appear in:} & \textbf{\color{ruby}Proceedings IEEE 24\textsuperscript{th}\,International Conference on Industrial Informatics, 2026}.\\
\ignore{\textbf{DOI:} & \textbf{\color{darkruby}10.1109/INDIN68935.2026.xxxxxxxx}\\}
& \\
\end{tabular}
}

\author{
  {\textbf{Raymond~Leung and Arman~Melkumyan}\vspace{2mm}}\\
  \begin{tabular}{c}
  Faculty of Engineering\\
  The University of Sydney\\
  \end{tabular}\vspace{2mm}
}

\renewcommand{\headerleft}{Effective information gathering and perspectives on adaptive sampling}
\renewcommand{\headeright}{Leung and Melkumyan, INDIN 2026}
\renewcommand{\undertitle}{Accepted manuscript}

\begin{document}
\maketitle

\begin{abstract}
A computational\,/\,analytics framework for assessing the value of drill-hole information in ore grade estimation is described using Gaussian Process and statistics. A distinguishing feature is that it presents both a near-term and long-term vision, circumvents conditional simulations and avoids making rigid assumptions such as stationarity and uncorrelated errors. Two experiments are devised to cater for situations where geological domains are differentiated or mixed. In scenario 1, performance (learning) curves are obtained to inform in-fill drilling and spacing consideration consistent with current practice. Analysis shows it is possible to estimate the incremental cost and reward via a proxy measure without relying on the {ground truth}, using insights obtained from a similar deposit, adjacent bench or domain. Scenario 2 examines adaptive sampling strategies and focuses on applying these in geologically complex areas with discontinuities and heterogeneous composition. Evaluation is made based on structural similarity, the mean and uncertainty in the posterior predictive distribution for the grade. The results highlight situations where regular grid sampling is suboptimal, and demonstrate an adaptive strategy that targets spatial complexity is capable of narrowing this gap. The proposed methodology can potentially be used in the future in an exploration--exploitation setting that involves sampling, machine learning, reasoning and cooperation between robots with embodied intelligence on a mine site.
\end{abstract}

\section*{Keywords}
Industrial informatics \sep data analytics \sep adaptive sampling \sep open-pit mining \sep mineral resource estimation.

\section*{MSC 2020 codes}
68U99 (Computing methodologies \& applications) \sep 60G15 (Gaussian processes) \sep\\ 65D15 (Algorithms for approximation of functions) \sep 62L05: (Sequential design of experiments).

\newpage
\section{Introduction}\label{sec:intro}
This study describes a multi-faceted approach to information gathering and examines how the sampling strategy (particularly, the density and location of measurements) would affect prediction performance for ore grade estimation in open-pit mining. For clarity, this paper is pivoted toward the informatics and computational intelligence\revisit{AI in industry} track of the conference and applications to the mineral resources industry. As background, the iron ore mining industry has invested heavily in mining automation and technology innovation to improve safety and productivity, as it commits to reduce carbon emission and embrace the transition to renewable energy \cite{holmes2022introduction,junior2024smart}. Some of the engineering and operational challenges\,/\,opportunities are explored in the survey article in \cite{leung2025automation}. One aspect worth noting is the breadth and interdisciplinary nature of this research, examples include mineral classification using hyperspectral imaging (autoencoder and CNN) \cite{windrim2023unsupervised}, online planning and vehicle dispatch algorithms \cite{seiler2020flow}, dynamic rail traffic optimization \cite{vujanic2022computationally}, portfolio\,/\,mining supply chain optimization \cite{zhou-samavati2021heuristics} just to name a few. Recent research and hardware development efforts also encompass the use of digital twins \cite{qu2023digital}, AI-driven optimization under uncertainty for mineral processing (formulated using partially observable Markov decision process) \cite{xu2025ai}, navigation and perception frameworks in a mine-site inspection robot \cite{liu2025dipper}. Whether the objective is to maximize profit or operational efficiency, simulation and probabilistic inference frameworks can only be as effective and reliable as the quantity and quality of the information they are served. Grade control and high-precision material tracking \cite{leung2025bucket} can both benefit directly from more informed knowledge of the deposit, which helps improve the accuracy of modeled boundaries that delineate ore and mine waste \cite{lowe2022bayesian}. Furthermore, guidance for semi-supervised data collection, short-term mine planning and stochastic production schedule optimization \cite{huang2020stochastic} could also benefit from minimizing the grade uncertainty. In this spirit, this study seeks to understand the cost and reward of additional sampling. It investigates the value of information associated with various drilling\,/\,sampling patterns under different conditions.

\begin{figure}[!htb]
\centerline{\includegraphics[width=0.6\columnwidth,trim={0 0 0 0},clip]{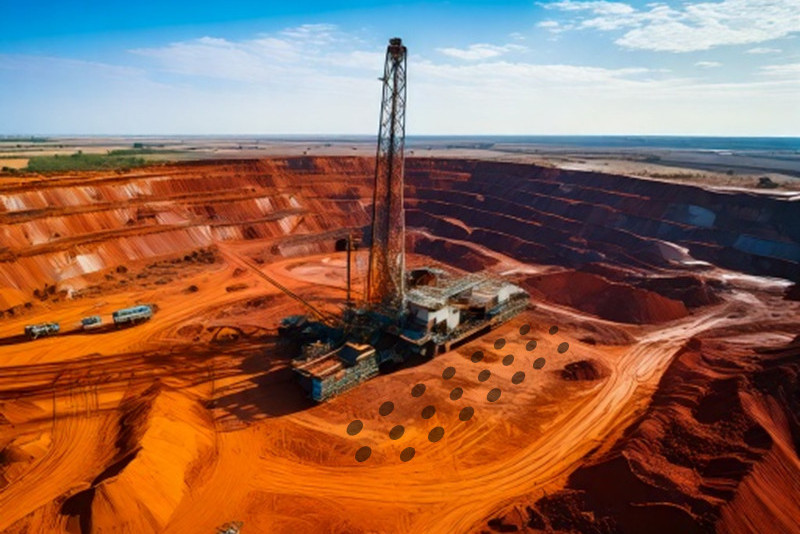}}
\caption{Open-pit mining environment with a typical drill pattern on a drilled bench. Disclaimer: Image is AI-generated using Adobe Firefly.}
\label{fig:operating-environment}
\end{figure}

\section{Background and Contributions}\label{sec:bkgd-contrib}
Unlike aerial geophysical surveys which rely on remote sensing, detailed knowledge of the subterranean geology is obtained mainly from drilling campaigns. This operation (depicted in Fig.~\ref{fig:operating-environment}) involves physically drilling into an ore deposit, taking samples from selected drilled holes, performing assay analysis either in the laboratory or with a portable XRF machine, to determine the geochemical composition (e.g. concentration of Fe). Traditionally, these processes are done manually, so it is sequential and slow as safety buffers need to be built-in for mining personnel. Even with efficiency improvement from automation, RC drilling and assaying can still cost \$30-100/m and \$5-50/sample. This roughly equates to \$1000 every 3 holes over a 10m bench. Typically, hundreds to thousands of blastholes are drilled during the mine production phase. Hence, there is a strong incentive to eliminate wasteful drilling that brings about little performance gain.

In current practice, the primary means to regulate cost is by adjusting the drill-hole density (or spacing) assuming a uniform drill pattern. During the exploration (coarse sampling) phase, this is guided by variogram-based heuristics.\footnote{In geostatistics \cite{oliver2015basic}, a variogram $\gamma(d)\propto\sigma^2_{\max}-\phi_{XX}(d)$ basically mirrors the autocovariance function, $\phi_{XX}(d)$, used in signal processing without assuming stationary second-order statistics. It describes the spatial dissimilarity between a pair of points, which is purely a function of the separating distance ($d$) for intrinsic stationary processes.} For instance, with spacing set to 2/3 of the variogram range. A legitimate criticism is that these rules are seldom justified or validated by evidence; often uncorrelated errors and stationary distributions are simply assumed \cite{taylor2019spacing}. Proposals that are grounded in geostatistics generally use conditional simulation (e.g. ordinary kriging with sequential Gaussian simulations) to estimate the impact on the uncertainty as new data is incorporated into the prediction model. This usually limits the scope of evaluation as repeated simulations can be computationally expensive. In Gomes et al., uniform drill-hole spacing optimization is driven by profit consideration \cite{gomes2025drill}. While this serves a useful purpose, we have chosen not to venture down this path in this study as other activities such as probabilistic material tracking would not benefit directly from end-to-end modeling and cost-revenue calculations. The dig limits described in \cite{gomes2025drill} also involve ore/waste classification which is sensitive to the cutoff grade \cite{chlingaryan2025integralgp}; so it is not compatible with preserving probabilistic predictions. In \cite{nowak2019optimal}, Nowak and Leuangthong assess the utility of drill-hole information based on a sufficiency test which determines whether the new measurements would likely reduce the uncertainty ($\pm z_{\alpha}\times \hat{\sigma}$ associated with a 90\% CI) enough to alter the geological confidence designation assigned to a patch.\footnote{For financial reporting (e.g. to ASX), a less than 15\% estimation error is commonly attached to the \textit{indicated} resource category, as opposed to \textit{inferred} resource where higher errors are tolerated \cite{lindi2024uncertainty}.}

This work addresses the sampling problem from a dual perspective. It contrasts regular spacing consideration under current practice with intelligent approaches that can operate more freely with fewer structural constraints. It complements previous studies with large-scale evaluations while offering both a short-term and long-term view. Its novelty lies in using multiple performance metrics to provide insights, targeting regions with different complexity, and evaluating the performance of various strategies (incl. adaptive sampling) in future application scenarios. The main contributions include
\begin{enumerate}
\item Performance curves that provide guidance for in-fill drilling or mine planning consistent with current practice;
\item Systematic comparison of drilling strategies in geologically complex areas consistent with autonomous robotic sampling which may be deployed in the future.
\end{enumerate}

\section{Conceptual Framework}\label{sec:conceptual-framework}
Our investigation comprises two sets of experiments which target different scenarios but they share a common workflow. In experiment 1, the objective is to dissect the multi-factorial performance metrics to reveal how predictive performance varies with sampling density. In experiment 2, the goal is to understand the effectiveness of various sampling strategies when applied to areas with high geological\,/\,structural complexity. The scenarios and metrics will be introduced after the computation steps are described. Throughout, the Fe predictive distribution is modeled using Gaussian Process (GP) regression \cite{leung2024eup3m-mdpi}. The general workflow contains five steps.
\begin{enumerate}
\item \textbf{Data synthesis} generates training sets $\mathcal{S}^\text{train}_{d+f}(z_\text{RL},g,\theta)$, test data and {ground truth} $\mathcal{S}^\text{test}(z_\text{RL},g)$ by sampling a high resolution model which describes the spatial distribution of iron at a real mine site. Without loss of generality, this data may be characterized by the elevation of the drilled bench ($z_\text{RL}$), geological domain ($g$), drill pattern orientation ($\theta$) and a spacing variable $s\!=\!d+f$, where $d\!\ge\!1$ and $f\in[0,1)$ denote integer and fractional density states. A data loader subsequently retrieves the data required for GP training.
\item \textbf{Learning} optimizes the Mat\'ern kernel hyperparameters $\vec{\omega}_{d+f}(z_\text{RL},g,\theta)$ used in GP regression by minimizing the negative log marginal likelihood \cite{chlingaryan2025integralgp}. This captures the spatial dependence between points using all measurements available in a training set, $\mathcal{D}\!\equiv\!\mathcal{S}^\text{train}_{d+f}(z_\text{RL},g,\theta)$.
\item \textbf{Inference} predicts the target attribute, viz. Fe grade $y(\vec{x}_*)$ at locations $\vec{x}_*$ in the test set $\mathcal{S}^\text{test}(z_\text{RL},g)\subset\mathbb{R}^k$, using the learned hyperparameters. Since GP Bayesian inference uses a GP as a prior distribution and expresses the posterior predictive density as a Gaussian over the entire function space, the posterior mean and covariance equations (\ref{eq:gp-posterior-mean})--(\ref{eq:gp-posterior-cov}) provide the average and uncertainty estimates at all locations $\mat{X}_*=[\vec{x}_{*1},\ldots,\vec{x}_{*m}]\in\mathbb{R}^{k\times m}$ conditioned on using the data in $\mathcal{D}$.
\begin{align}
\vec{\mu}_*(\mat{X}_*\!\mid\!\mathcal{D}) &= \mathbf{\mu}(\mat{X}_*)+\mat{K}_*^T \mat{K}_y^{-1}(\vec{y}-\mathbf{\mu}(\mat{X}_*))\label{eq:gp-posterior-mean}\\
\mat{\Sigma}_*(\mat{X}_*\!\mid\!\mathcal{D}) &= \mat{K}_{**}-\mat{K}_*^T \mat{K}_y^{-1} \mat{K}_*\label{eq:gp-posterior-cov}
\end{align}
Further details on the $\mat{K}$ matrices can be found in \cite{chlingaryan2025integralgp}.
\item \textbf{Evaluation} considers which location to sample next to expand the training data set $\mathcal{D}$ based on some criteria to be described in Sec.~\ref{sec:methods}. This applies only to strategies in experiment 2 which generally select points in a sequential manner. It is bypassed in experiment 1 as points are typically selected in batch in a deterministic way.
\item \textbf{Analysis} computes the metrics and compares the predictions (\ref{eq:gp-posterior-mean})--(\ref{eq:gp-posterior-cov}) with reference values (the {ground truth}) to establish performance as a function of the data used.
\end{enumerate}

\section{Experiments: Scenarios and Visions}\label{sec:experiments-scenarios}
Experiments were devised to cover two scenarios. Their key distinctions are summarized in Table~\ref{tab:scenarios}. Scenario 1 covers a large section of the Hamersley\,/\,Fortescue group deposit \cite{thorne2014structural}. Because the geological domains are differentiated---the modeled space is partitioned into mineralized, hydrated or waste domains, and the GP hyperparameters are learned on a per geozone basis---geological complexity is kept low, and the target function (observed Fe grade) tends to be more homogeneous within each domain. A hierarchical multi-resolution structure is used to control drill-hole spacing. As it produces an embedded and uniform drill pattern, this design is consistent with current industry practice in relation to in-fill drilling and spacing consideration. Scenario 2 focuses on sub-volumes with high geological complexity. To emulate a limited knowledge of  stratigraphic boundaries in a local region, here, the domains are not differentiated. The sub-volumes are chosen at locations where grade discontinuities, faults and igneous intrusion (complicated features) are found, to present challenging test cases for various sampling strategies. This vision is created in anticipation that advanced robotics with embodied intelligence could one day be deployed to collect and analyze rock samples with minimal delay, and perhaps even incorporate collaborative learning and cooperative exploration-exploitation strategies to optimize mining processes and operate autonomously at scale; enabled by AI and distributed (or edge) computing.

\begin{table}[!htb]
\caption{Scenarios corresponding to the two experiments}
\centering
\begin{tabular}{|l|l|l|}
\hline
& Scenario\,/\,Experiment 1 & Scenario\,/\,Experiment 2\\\hline
Spatial scale & large (approx.\,1.6$\times$2.5 km) & local (max. 300-500m)\\
Structure & low complexity & high complexity\\
Domains & differentiated (14 individual) & undifferentiated (mixed)\\
Distribution & Fe grade is mostly unimodal & multimodal\\
Drill pattern & embedded, regular & unconstrained, irregular\\
Deployment & in line with current practice & with autonomous agents\\\hline
\end{tabular}
\label{tab:scenarios}
\end{table}

\begin{figure}[!htb]
\centerline{\includegraphics[width=0.6\columnwidth]{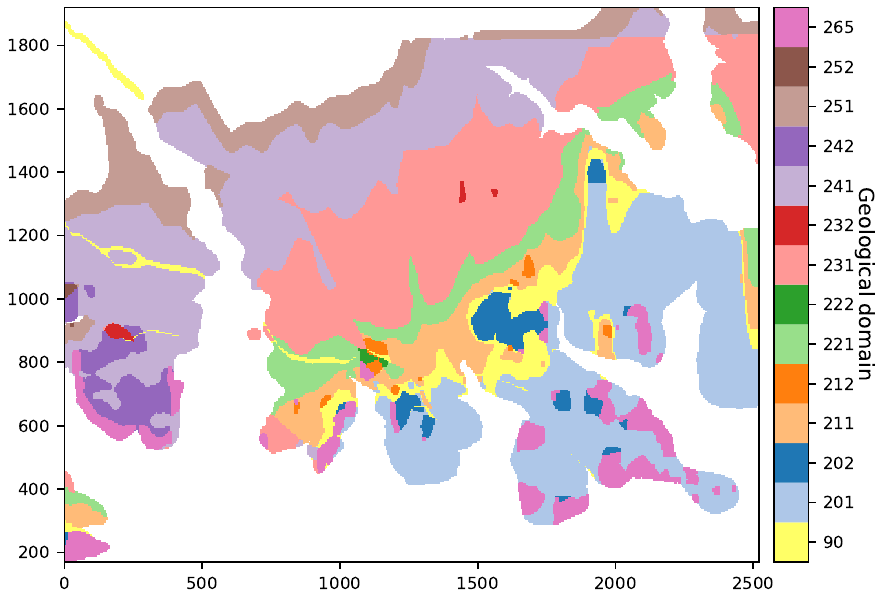}}
\caption{Domain structure of the site (pit) considered in this study. It has an area of 1600$\times$2500\,m\textsuperscript{2} and z-interval of 80m. Resolution is $5\times 5\times 2.5$m. Only bench 560 [$z_\text{RL}\in[560,570)$] is shown. Dolerite, waste, mineralized and hydrated domains are assigned a last digit of 0, 1, 2 and 5, respectively.}
\label{fig-site-domains-560}
\end{figure}
\section{Methods and Implementation}\label{sec:methods}
\begin{figure}[!htb]
\centerline{\includegraphics[width=0.6\columnwidth]{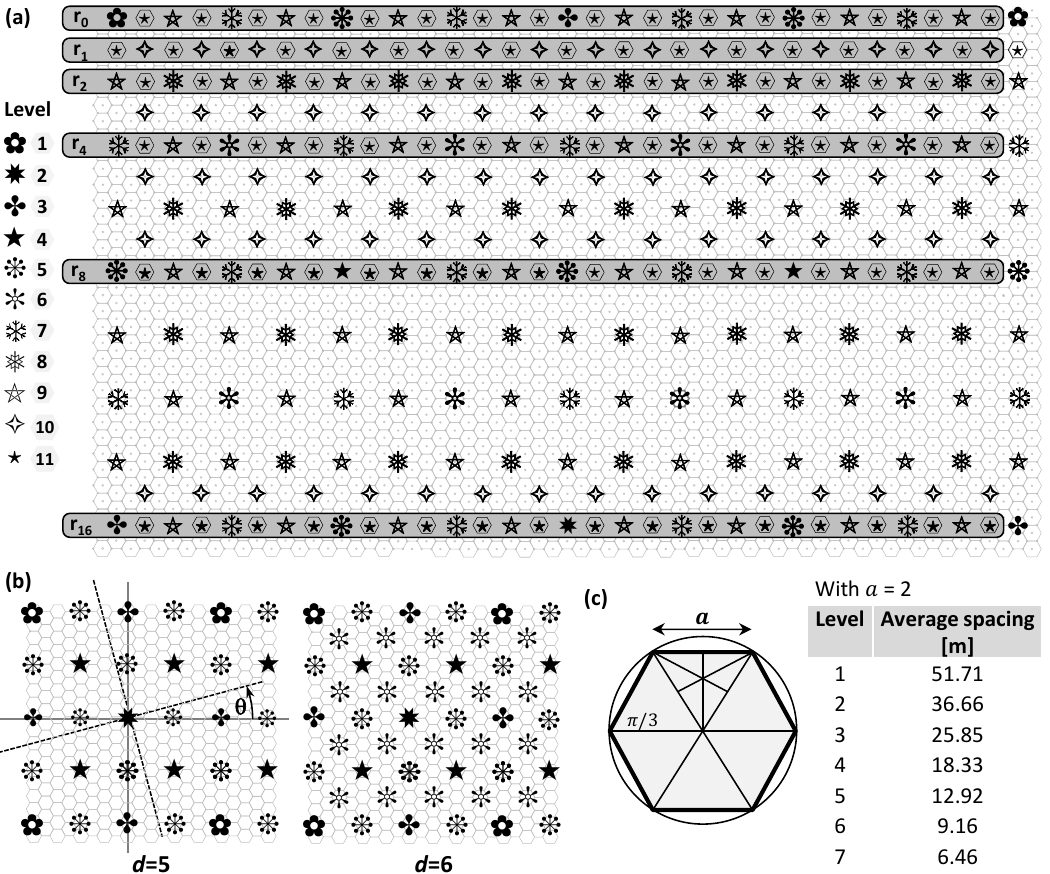}} 
\caption{Sampling on a hexagonal lattice. (a) Spatial hierarchy with $d\!=\!1$ being the coarsest, $d\!=\!D$ being the densest. The $2^{(D-1)/2}$ periodic pattern can be constructed using labels in vectors $\vec{r}_0$ and $\vec{r}_{2^k}, k\in\mathbb{Z}\le (D-3)/2$. (b) Training points at level 5 and 6. (c) Average spacing with $a\!=\!2$.}
\label{fig-multires-hexgrid}\vspace{3mm}
\end{figure}

The domain structure in Fig.~\ref{fig-site-domains-560} represents the high \nobreak{resolution} data source from which sampling takes place. This synthesis approach gives complete freedom to how sampling is conducted and perfect knowledge of the true grades for a prototypical deposit. It represents a departure from the status quo where modelers usually attempt to replicate the {ground truth} through simulations, often with limited reliability given insufficient data. Scenario 1 synthesizes data sets by picking points on a hexagonal lattice superimposed onto this structure. Fig.~\ref{fig-multires-hexgrid}(a) shows this lattice and identifies the points $\mathcal{P}_d$ that belong to each level $d$ in the spatial hierarchy. In Fig.~\ref{fig-multires-hexgrid}(b), the embedded property ensures the training set at level $d$ is the result of combining the current and all previous subsets, viz., $S_d=\cup_{i=1}^{d}\mathcal{P}_d$; this yields a recursive decomposition with rectangular\,/\,diamond point constellation. In Fig.~\ref{fig-multires-hexgrid}(c), we set the scaling parameter $a\!=\!2$ to provide meaningful coverage of average drill-hole distances roughly in the 6m to 50m range. In between levels, fractional increment [the addition of $f\times(\lvert S_{d+1}\rvert - \lvert S_d\rvert)$ points from $\mathcal{P}_{d+1}$] is conducted with first preference given to points closest to domain boundaries.

The \textbf{sampling strategies} employed in scenario 2 may be viewed as sequential algorithms that are unencumbered by preconceived notions or present day constraints. At each iteration, with $n$ drill-hole samples already acquired under policy $\pi$ in training set $\mathcal{D}^{\pi}_n$, the learning and inference steps are performed to (i) obtain updated kernel hyperparameters $\vec{\omega}^{\pi}_n(z_\text{RL},g)$ using $\mathcal{D}^{\pi}_n$; (ii) obtain posterior mean and standard deviation estimates, $\hat{\mu}^{\pi}_n(\vec{x}_*)$ and $\hat{\sigma}^{\pi}_n(\vec{x}_*)$ for $\vec{x}_*\in S^\text{test},z_*\in[z_\text{RL},z_\text{RL}\!+\!20)$ using $\vec{\omega}^{\pi}_n(z_\text{RL},g)$. The goal is to select a location $\tilde{\vec{x}}$ to collect the next set of measurements down a drill-hole according to policy $\pi$. The adaptive strategies consist of MaxVariance (MV), TargetFeature (TF), TargetComplexity (TC) and UncertaintyReduction (UR). Random selection (RN) and rectangular Grid layout (GD) are also included for comparison.

\begin{itemize}
\item MV sets $\tilde{\vec{x}}\!=\!\argmax_{\vec{x}_*} \hat{\sigma}^\text{\tiny{MV}}_n(\vec{x}_*)^2$ and seeks out locations with the highest epistemic uncertainty (i.e. points furthest away from past measurements).
\item TF sets $\tilde{\vec{x}}\!=\!\argmax_{\vec{x}_*} \hat{\sigma}^\text{\tiny{TF}}_n(\vec{x}_*)[1+\eta\cdot g(\hat{\mu}^\text{\tiny{TF}}_n(\vec{x}_*))]$ where $g(\hat{\mu}^\text{\tiny{TF}}_n(\vec{x}_*))$ represents a gradient-based prior with values standardized to $[0,1]$. This is computed using a derivative operator (3D Sobel filter) which is applied to the posterior mean [a length-scale and sampling state dependent GP smoothed random field]. The idea is to accentuate geological features of interest \cite{hwang2019auv} and use this to influence $\hat{\sigma}^\text{\tiny{TF}}_n$ based point selection. To balance these dynamics, the tuning parameter ($\eta$) is set to $\mathbb{E}[(\hat{\vec{\mu}}^\text{\tiny{TF}}_n - \mathbb{E}[\hat{\vec{\mu}}^\text{\tiny{TF}}_n])^2]^{-1/2}$.
\item TC sets $\tilde{\vec{x}}\!=\!\argmax_{\vec{x}_*} \hat{\sigma}^\text{\tiny{TC}}_n(\vec{x}_*)[1+h(\vec{x}_*)]$ where $h(\vec{x}_*)\!=\!\text{Var}[Y(\vec{x})]^{1/2}/\mathbb{E}[Y(\vec{x})]$ for $\vec{x}\in\mathcal{N}(\vec{x}_*)$ represents a distance-weighted coefficient of variation in the grade, computed using $K$ training points from $\mathcal{D}^\text{\tiny TC}_n$ in the nearest neighborhood of $\vec{x}_*$. This formulation emphasizes areas with higher local signal variance.
\item UR sets $\tilde{\vec{x}}\!=\!\argmax_{\vec{x}_*} \mathbb{E}[\hat{\sigma}^\text{\tiny{UR}}_n(X_*\!\mid\!\mathcal{D}^\text{\tiny UR}_n\cup\{\vec{x}_*\})^2 - \hat{\sigma}^\text{\tiny{UR}}_n(X_*\!\mid\!\mathcal{D}^\text{\tiny UR}_n)^2]$, where the expectation is taken over all $m$ candidate locations in $\mat{X}_*$. Essentially, it chooses the point that maximizes uncertainty reduction.
\end{itemize}
A full implementation might interpret $[\,\cdot\!\mid\!\mathcal{D}^\text{\tiny UR}_n\cup\{\vec{x}_*\}]$ as requiring conditional simulations, drawing $L$ simulated grade values $\tilde{y}^{(l)}_n$ from $N(\hat{\mu}^\text{\tiny{UR}}_n(\vec{x}_*),\hat{\sigma}^\text{\tiny{UR}}_n(\vec{x}_*))$, then re-learning the hyperparameters $\tilde{\vec{\omega}}^{(l)}_{n*}$ for each random realization, before $\hat{\sigma}^\text{\tiny{UR}}_n(X_*\!\mid\!\mathcal{D}^\text{\tiny{UR}}_n\cup\{\vec{x}^{(l)}_n\})\equiv\hat{\sigma}^\text{\tiny{UR}}_n(X_*\!\mid\!\tilde{\vec{\omega}}^{(l)}_{n*})$ is estimated for points in $\mat{X}_*$. Since the GP variance is purely a function of points geometry and the kernel (i.e. completely independent of $y$ values), we argue the simulation makes little difference to the average provided the hyperparameters are stable w.r.t. small perturbance. In practice, it suffices to estimate $\hat{\sigma}^\text{\tiny{UR}}_n(X_*\!\mid\!\mathcal{D}^\text{\tiny{UR}}_n\cup\{\vec{x}^{(l)}_n\})$ using $\vec{\omega}^\text{\tiny{UR}}_n$. The candidates for argmax consideration, $\vec{x}_*$, may be limited to the top $M\propto\sqrt{m}$ (say, 30) with highest average $ \hat{\sigma}^\text{\tiny{TC}}_n(\vec{x}_*)$ value over the drill-hole. Furthermore, the difference calculation can be confined to points in $\mat{X}_*$ within the sphere of influence of the candidate location, $\vec{x}_*$, as informed by the kernel length-scale parameters.

The \textbf{performance metrics} used in scenario 1 cover global distortion, spatial fidelity and prediction uncertainty. This includes Jensen-Shannon divergence (JSD), an information-theoretic measure of the differences between model prediction and true grade histograms. To account for spatial correlation, a fidelity statistic (F) is used to measure quality degradation in terms of the median ratio between the predicted and true grade variograms. To quantify uncertainty, the interval tightness statistic (I) is used along side the coefficient of variation (CV). The definition of JSD, F and I are given in \cite{leung2024eup3m-mdpi}. For scenario 2, since the random process and error statistics are generally non-stationary, more localized (pixel-based) measures are used. This includes the RMSE, the structural similarity index (SSIM), and continuous ranked probability score (CRPS) which compares the predictive distribution with the true value. SSIM and CRPS are specified in \cite{wang2004image} and \cite{gneiting2005calibrated}.

\section{Results and Analysis}\label{sec:results}
\subsection{Experiment 1}\label{sec:results-exp1}
\begin{figure}[!thb]
\centerline{\includegraphics[width=0.6\columnwidth]{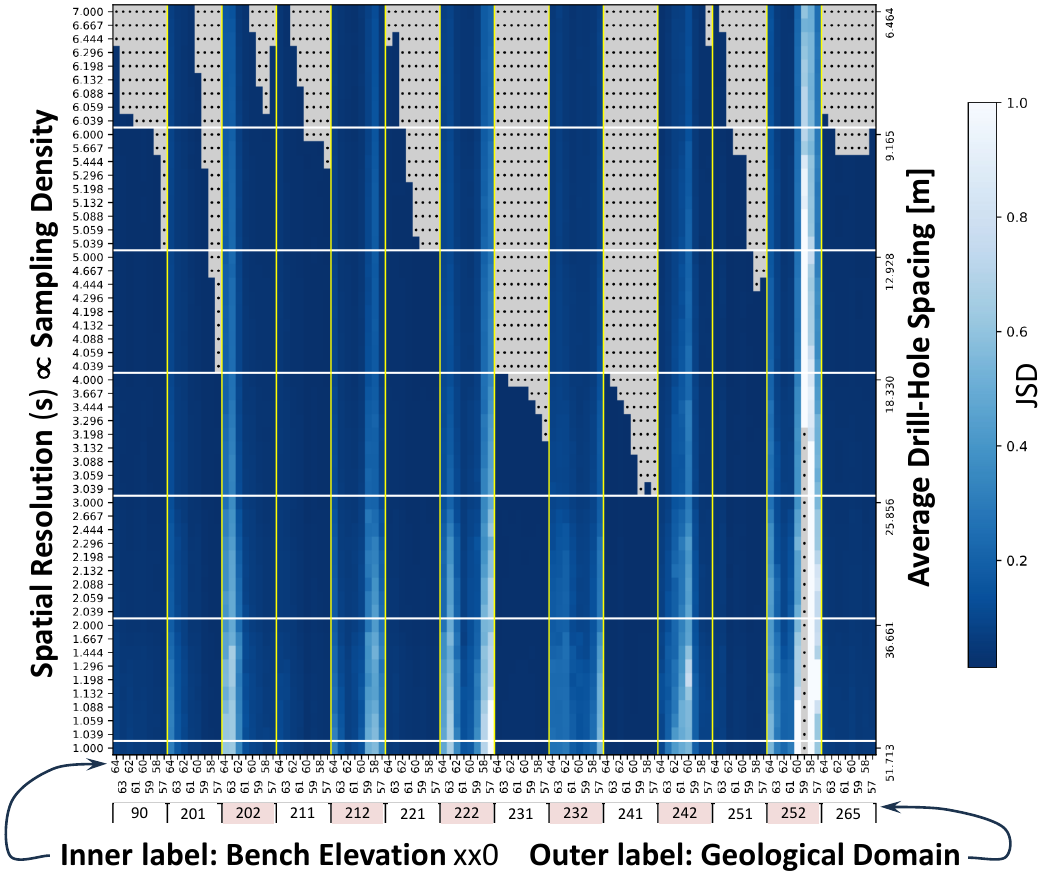}} 
\caption{Image-based representation of raw JSD scores averaged over different drill pattern orientations $\theta\in\{0,\frac{\pi}{6},\frac{\pi}{4},\frac{\pi}{3},\frac{\pi}{2},\frac{2\pi}{3},\frac{3\pi}{4},\frac{5\pi}{6}\}$.}
\label{fig-statimage-interpretation}
\end{figure}

To appreciate the scale of the experiments and understand how the results are interpreted, Fig.~\ref{fig-statimage-interpretation} renders one of the computed metrics $\zeta_{m,s}^{(g,z_\text{RL})}$ as an image for $m=\text{JSD}$. The left vertical axis represents spatial resolution (level $s\equiv d+f$) and the sampling density increases upward from $s\!=\!1$ to $s\!=\!D$.
The right axis provides an equivalent description in terms of average drill-hole spacing. Data is sparse at the bottom and the spacing shrinks by $\sim\!70\%$ as we climb past each horizontal grid line. A blue colour scale is used for argmin objectives (distortion\,/\,uncertainty metrics such as JSD). For argmax objectives (fidelity metric F), a red colour scale is used and the intensity is flipped vertically to ensure a darker shade always indicates better performance. The horizontal axis specifies the modeled region. It groups results by geozones ($g$) and interleaves bench elevation ($z_\text{RL}$) within each group. Dotted pixels denote empty configurations where either training points are unavailable (usually when sampling at low density) or omitted when the sample size has reached the upper limit, with $\lvert S^\text{train}_{d+f}\rvert\ge 3200$. The main take-away is that each domain has its own goal post for the absolute performance that can be achieved, as evident from the variation in intensity across the domains. Mineralized domains (xx2) tend to be brighter, hence, more challenging to model.

To make changes w.r.t. sampling more conspicuous, raw metrics are normalized in each column according to (\ref{eq:normalized-metric}).
\begin{align}
\xi_{m,s}^{(g,z_\text{RL})}=\frac{\zeta_{m,s}^{(g,z_\text{RL})}}{\kappa},\,\,\kappa\defeq\begin{cases}\max_s \zeta_{m,s}^{(g,z_\text{RL})}\!&\!\text{for fidelity}\\ \min_s \zeta_{m,s}^{(g,z_\text{RL})}\!&\!\text{otherwise}\end{cases}
\label{eq:normalized-metric}
\end{align}
\begin{figure*}[!tbh]
\centering
\resizebox{\textwidth}{!}{
\setlength{\tabcolsep}{1pt}
\begin{tabular}{ccc}
\includegraphics[width=0.33\textwidth]{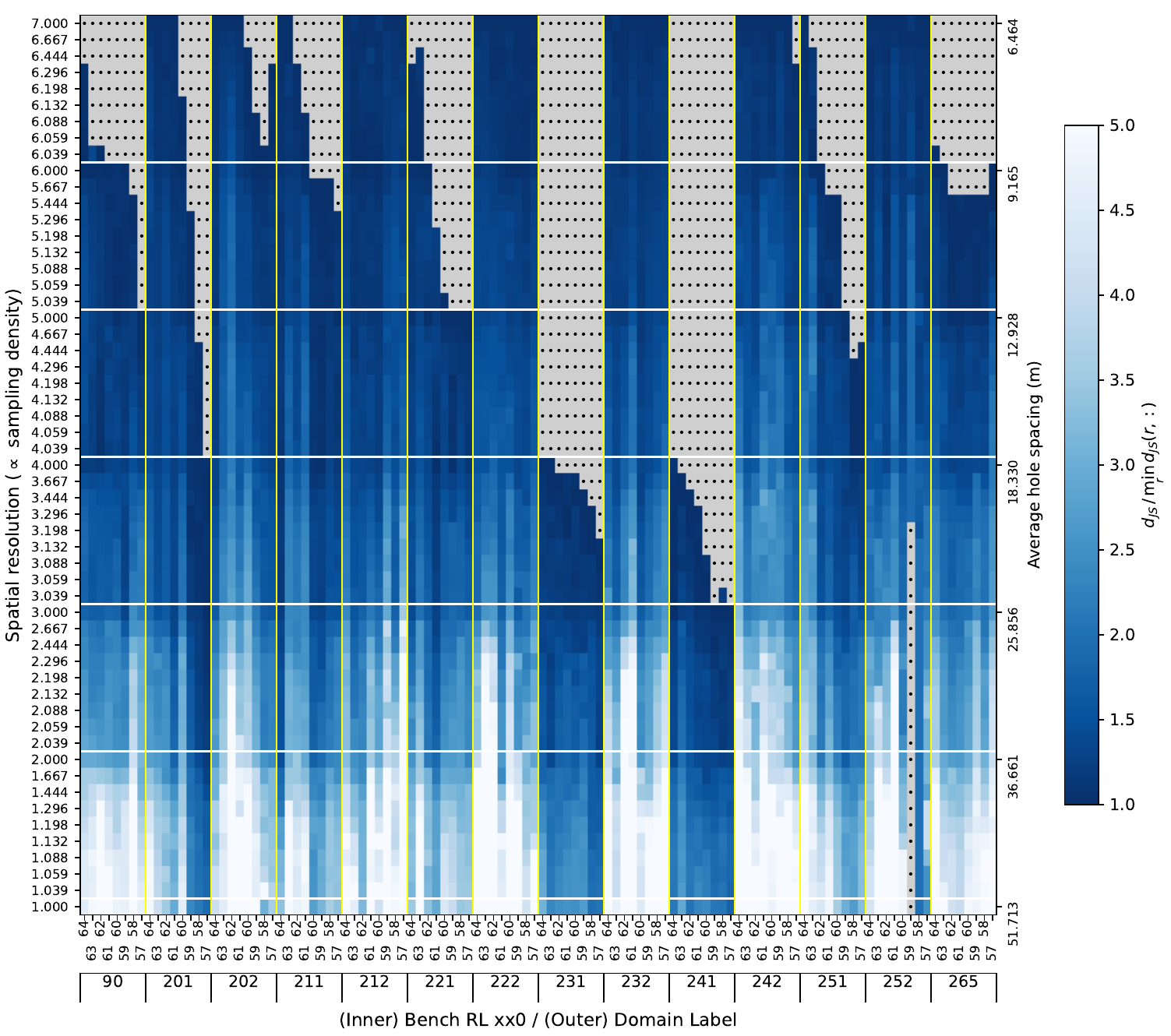} &
\includegraphics[width=0.33\textwidth]{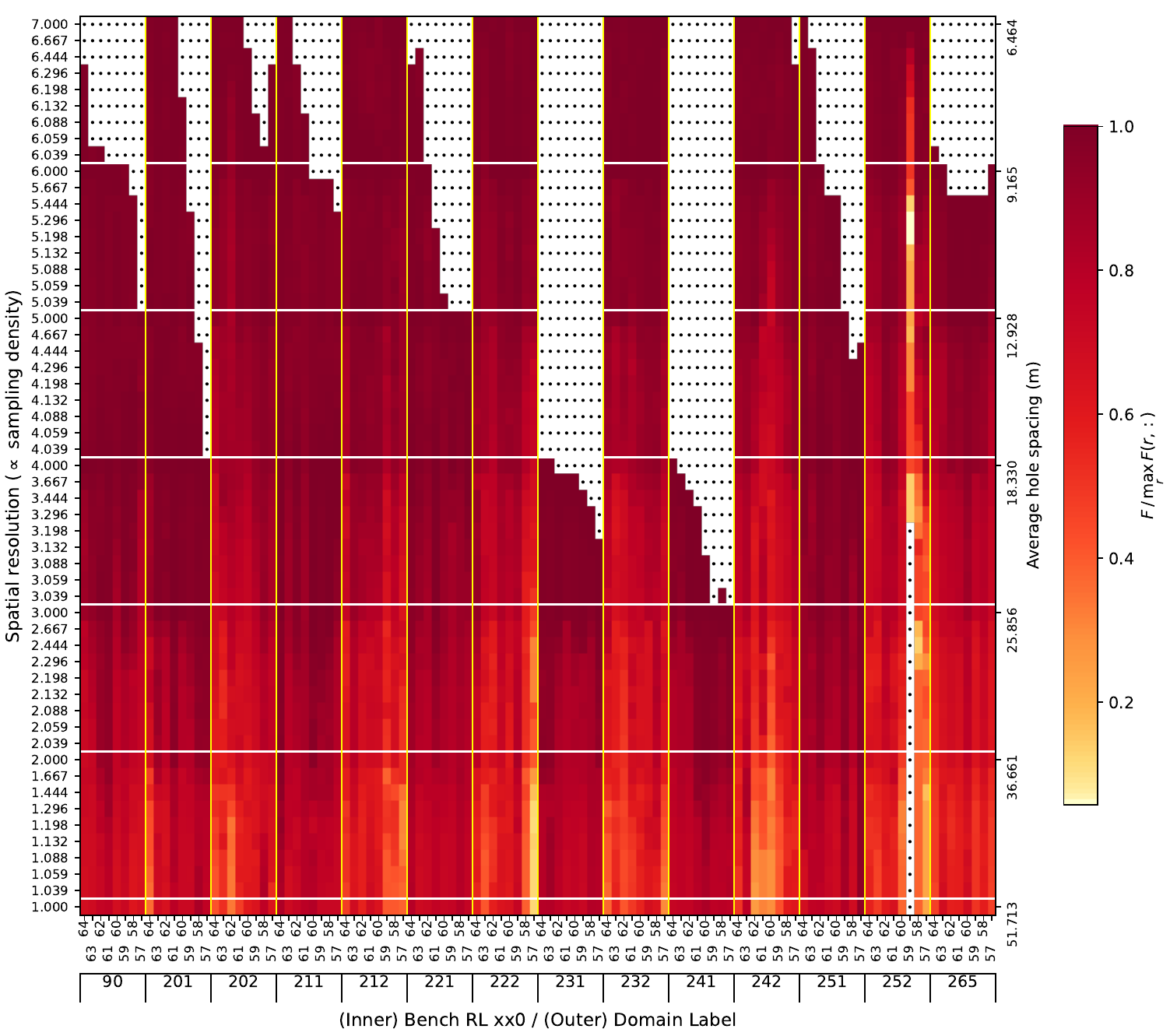} &
\includegraphics[width=0.33\textwidth]{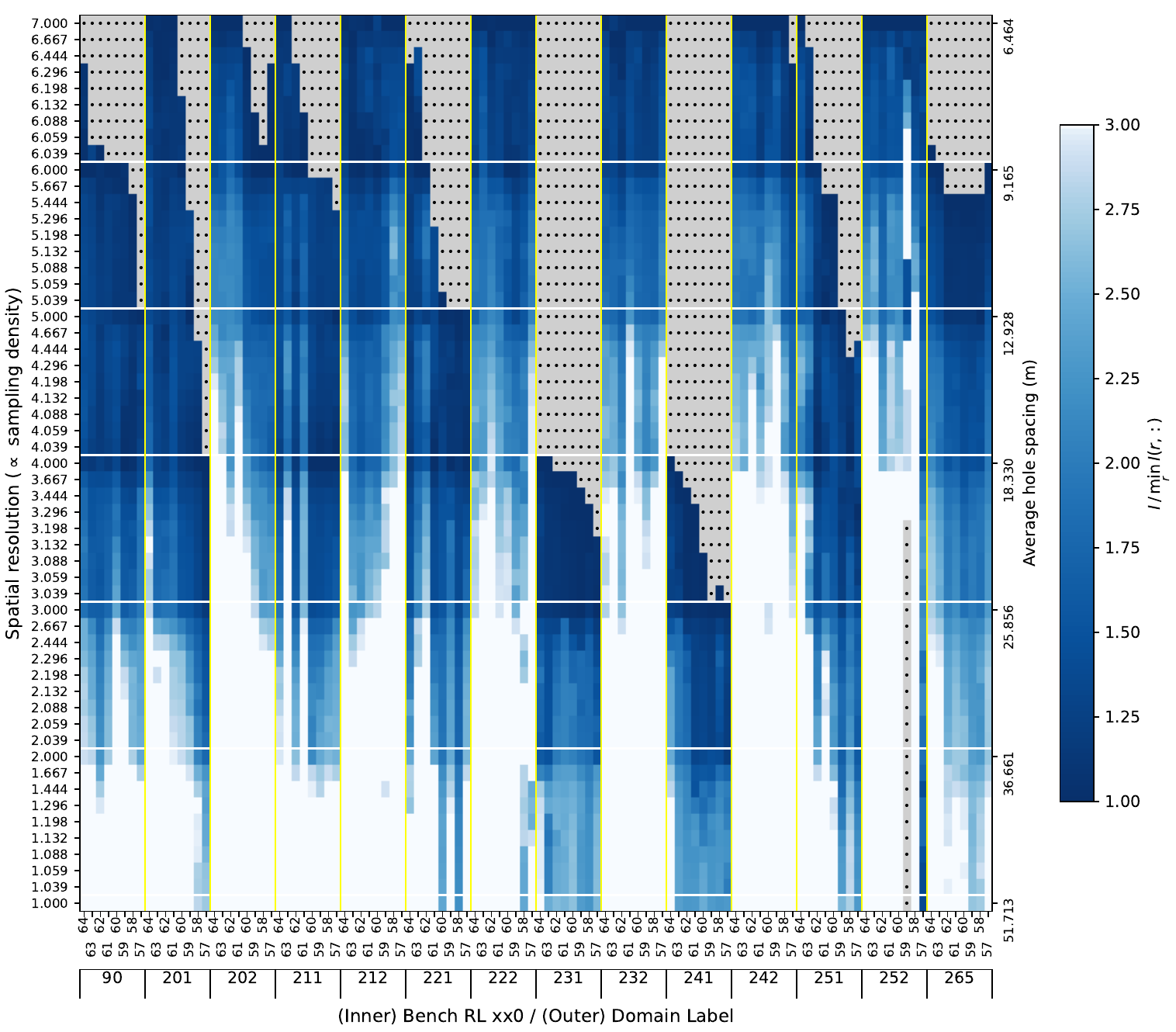}
\end{tabular}
}
\caption{Image of normalized metrics $\xi_{m,s}^{(g,z_\text{RL})}$. Left to Right: global distortion (JSD), spatial fidelity (F), interval tightness (I)}\label{fig:normalized-images}
\end{figure*}
\begin{figure*}[!htb]
\resizebox{0.95\textwidth}{!}{
\setlength{\tabcolsep}{1pt}
\begin{tabular}{ccc}
\includegraphics[width=0.32\textwidth]{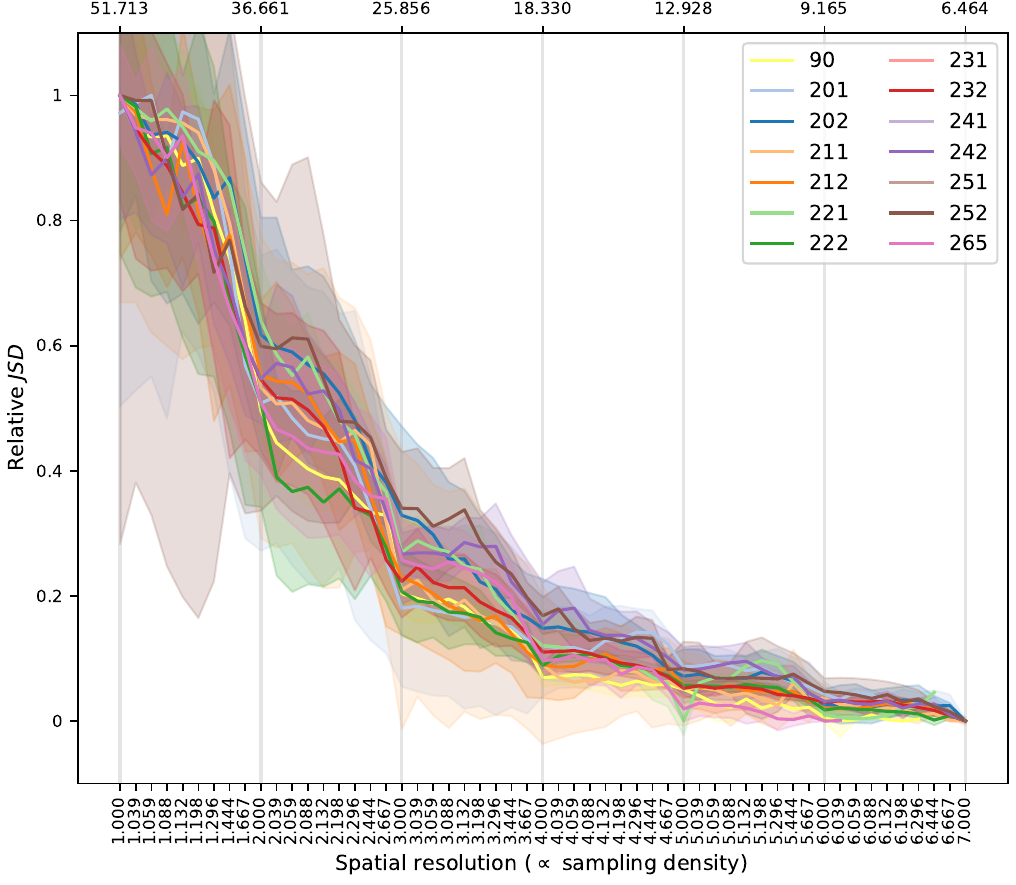} &
\includegraphics[width=0.32\textwidth]{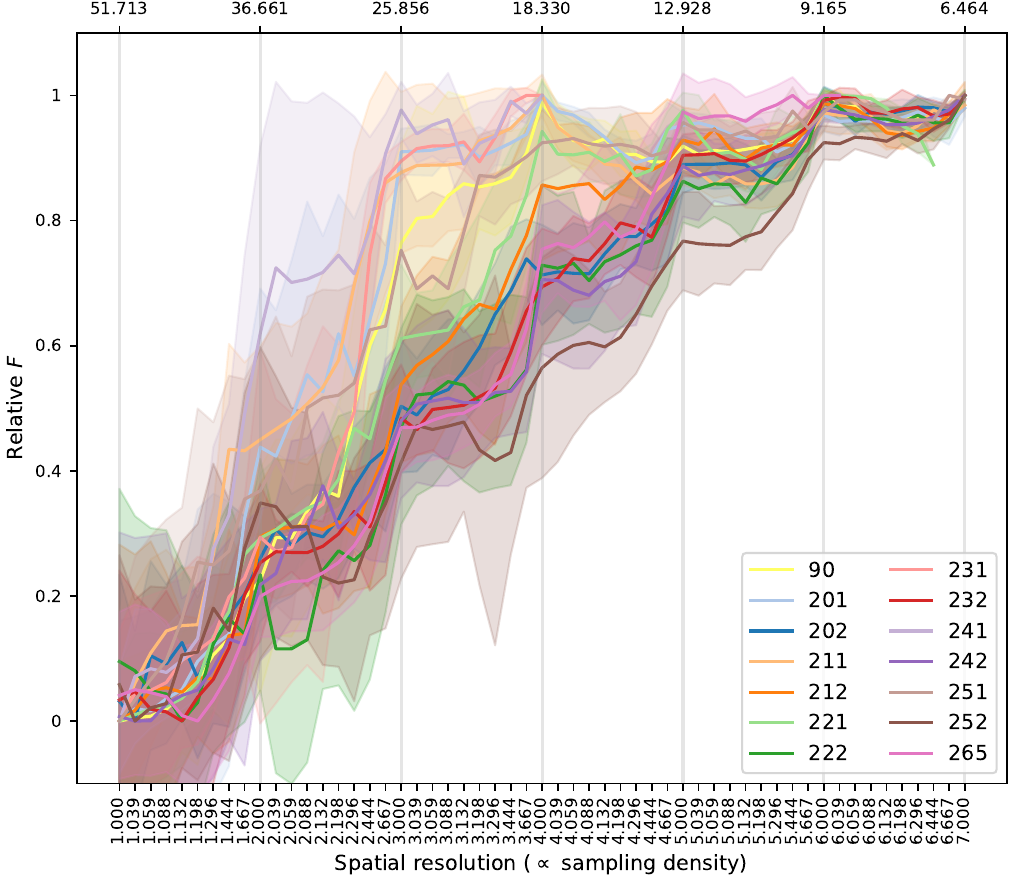} &
\includegraphics[width=0.32\textwidth]{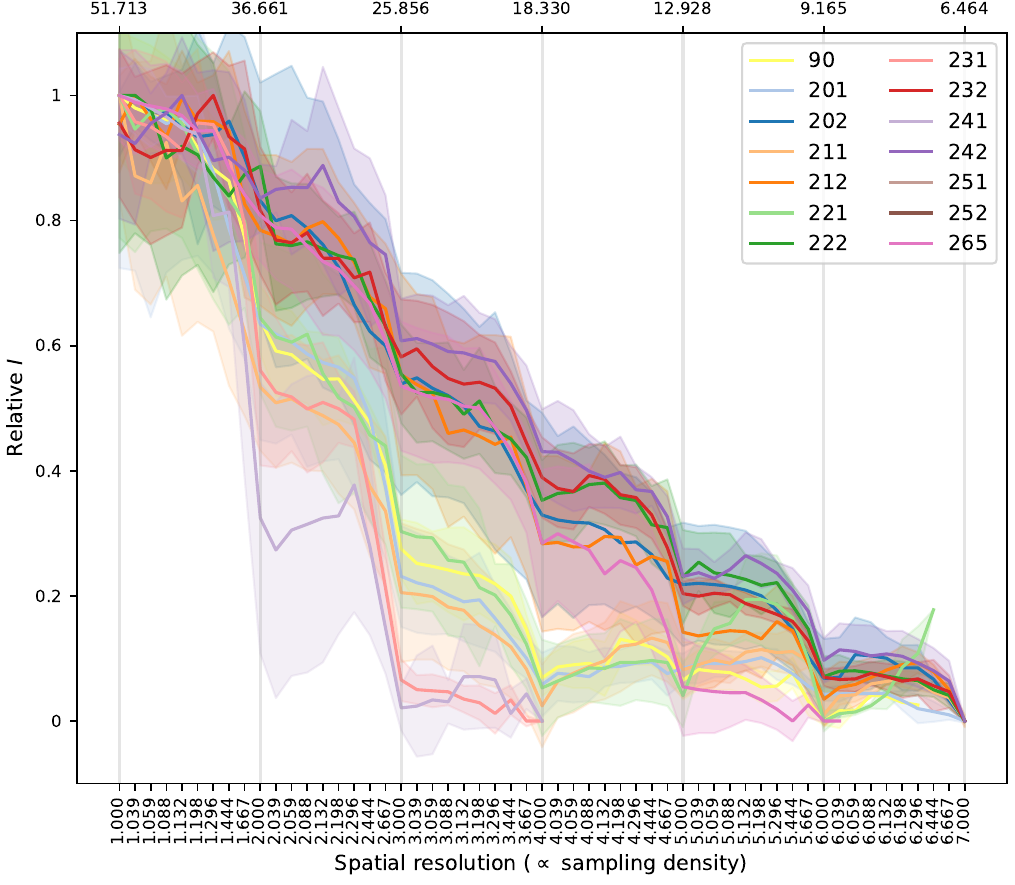}\\
\end{tabular}
}
\caption{Performance curves for standardized metrics $\check{\xi}_{m,s}^{(g)}$. Left to Right: global distortion (JSD), spatial fidelity (F), interval tightness (I)}\label{fig:individual-performance-curves}
\end{figure*}
The increased contrast in Fig.~\ref{fig:normalized-images} shows the normalized metrics (JSD, F and I) clearly respond to changes in sampling density. In general, the range of improvement is narrower for waste domains (xx1). Conversely, the relative gain is higher for mineralized\,/\,hydrated (xx2\,/\,xx5) domains. This finding vindicates the general practice of increasing focus and drilling in mineralized and hydrated domains. Comparing JSD with the prediction interval tightness statistic in Fig.~\ref{fig:normalized-images}, we see the potential for distortion reduction is higher; JSD can shrink by a factor of 5 or more. For uncertainty reduction, the transition from white to light blue occurs much later---the required spacing is around 25.85 m for JSD, but is roughly 18.3 m for interval tightness. Thus, the sampling requirement associated with uncertainty is higher than that warranted by global accuracy---requiring a full level increment or a doubling of training points to achieve essentially the same gain (i.e., reducing JSD and I both by x\%). Based on the observed $\xi_{\max}/\xi_{\min}$ ratios, the potential gain for each geozone and metric are reported in Table~\ref{tab:potential-gains}.
\begin{table}[!h]
\centering
\renewcommand{\arraystretch}{1.05}
\setlength\tabcolsep{6pt}
\caption{Performance gain factors observed in different domains, $G^{(g)}_m$}\label{tab:potential-gains}
\resizebox{0.5\columnwidth}{!}{
\begin{tabular}{|l|cccccc|}\hline
Metric, $m$ & 90 & 201 & 211 & 221 & 231 & 241\\\hline
\quad JSD & \cellcolor[rgb]{0.0314,0.3060,0.5944}{\color[rgb]{1,1,1}\textbf{5.832}} & \cellcolor[rgb]{0.4353,0.6910,0.8426}{\color[rgb]{0,0,0}\textbf{3.228}} & \cellcolor[rgb]{0.3313,0.6221,0.8048}{\color[rgb]{0,0,0}\textbf{3.732}} & \cellcolor[rgb]{0.3364,0.6255,0.8067}{\color[rgb]{0,0,0}\textbf{3.719}} & \cellcolor[rgb]{0.6110,0.7874,0.8805}{\color[rgb]{0,0,0}\textbf{2.506}} & \cellcolor[rgb]{0.6670,0.8123,0.8989}{\color[rgb]{0,0,0}\textbf{2.225}}\\
\quad F & \cellcolor[rgb]{0.7964,0.8721,0.9439}{\color[rgb]{0,0,0}\textbf{1.485}} & \cellcolor[rgb]{0.8023,0.8760,0.9459}{\color[rgb]{0,0,0}\textbf{1.418}} & \cellcolor[rgb]{0.8082,0.8800,0.9478}{\color[rgb]{0,0,0}\textbf{1.384}} & \cellcolor[rgb]{0.8112,0.8820,0.9488}{\color[rgb]{0,0,0}\textbf{1.357}} & \cellcolor[rgb]{0.8171,0.8859,0.9508}{\color[rgb]{0,0,0}\textbf{1.300}} & \cellcolor[rgb]{0.8141,0.8839,0.9498}{\color[rgb]{0,0,0}\textbf{1.312}}\\
\quad I & \cellcolor[rgb]{0.2658,0.5773,0.7792}{\color[rgb]{0,0,0}\textbf{4.067}} & \cellcolor[rgb]{0.2809,0.5876,0.7851}{\color[rgb]{0,0,0}\textbf{3.988}} & \cellcolor[rgb]{0.2522,0.5660,0.7731}{\color[rgb]{1,1,1}\textbf{4.150}} & \cellcolor[rgb]{0.3515,0.6358,0.8126}{\color[rgb]{0,0,0}\textbf{3.636}} & \cellcolor[rgb]{0.5922,0.7771,0.8764}{\color[rgb]{0,0,0}\textbf{2.584}} & \cellcolor[rgb]{0.6374,0.7997,0.8886}{\color[rgb]{0,0,0}\textbf{2.387}}\\
\quad CV & \cellcolor[rgb]{0.4171,0.6806,0.8382}{\color[rgb]{0,0,0}\textbf{3.295}} & \cellcolor[rgb]{0.3364,0.6255,0.8067}{\color[rgb]{0,0,0}\textbf{3.725}} & \cellcolor[rgb]{0.2197,0.5335,0.7563}{\color[rgb]{1,1,1}\textbf{4.366}} & \cellcolor[rgb]{0.3969,0.6669,0.8304}{\color[rgb]{0,0,0}\textbf{3.394}} & \cellcolor[rgb]{0.5671,0.7633,0.8710}{\color[rgb]{0,0,0}\textbf{2.688}} & \cellcolor[rgb]{0.6325,0.7976,0.8869}{\color[rgb]{0,0,0}\textbf{2.415}}\\
\hline\hline
Metric, $m$ & 265 & 202 & 212 & 222 & 232 & 242\\\hline
\quad JSD & \cellcolor[rgb]{0.1304,0.4442,0.7103}{\color[rgb]{1,1,1}\textbf{4.923}} & \cellcolor[rgb]{0.1710,0.4848,0.7312}{\color[rgb]{1,1,1}\textbf{4.654}} & \cellcolor[rgb]{0.1025,0.4087,0.6829}{\color[rgb]{1,1,1}\textbf{5.141}} & \cellcolor[rgb]{0.0314,0.2776,0.5522}{\color[rgb]{1,1,1}\textbf{5.990}} & \cellcolor[rgb]{0.0314,0.2126,0.4558}{\color[rgb]{1,1,1}\textbf{6.419}} & \cellcolor[rgb]{0.0314,0.1882,0.4196}{\color[rgb]{1,1,1}\textbf{6.552}}\\
\quad F & \cellcolor[rgb]{0.7309,0.8395,0.9213}{\color[rgb]{0,0,0}\textbf{1.888}} & \cellcolor[rgb]{0.7408,0.8437,0.9248}{\color[rgb]{0,0,0}\textbf{1.845}} & \cellcolor[rgb]{0.7408,0.8437,0.9248}{\color[rgb]{0,0,0}\textbf{1.835}} & \cellcolor[rgb]{0.6768,0.8165,0.9024}{\color[rgb]{0,0,0}\textbf{2.170}} & \cellcolor[rgb]{0.6867,0.8207,0.9058}{\color[rgb]{0,0,0}\textbf{2.109}} & \cellcolor[rgb]{0.6867,0.8207,0.9058}{\color[rgb]{0,0,0}\textbf{2.132}}\\
\quad I &  \cellcolor[rgb]{0.4729,0.7116,0.8507}{\color[rgb]{0,0,0}\textbf{3.082}} & \cellcolor[rgb]{0.2076,0.5213,0.7501}{\color[rgb]{1,1,1}\textbf{4.436}} & \cellcolor[rgb]{0.1179,0.4284,0.6983}{\color[rgb]{1,1,1}\textbf{5.030}} & \cellcolor[rgb]{0.1467,0.4604,0.7187}{\color[rgb]{1,1,1}\textbf{4.821}} & \cellcolor[rgb]{0.1117,0.4205,0.6921}{\color[rgb]{1,1,1}\textbf{5.072}} & \cellcolor[rgb]{0.1271,0.4402,0.7075}{\color[rgb]{1,1,1}\textbf{4.957}}\\
\quad CV & \cellcolor[rgb]{0.5420,0.7495,0.8656}{\color[rgb]{0,0,0}\textbf{2.798}} & \cellcolor[rgb]{0.2758,0.5842,0.7831}{\color[rgb]{0,0,0}\textbf{4.020}} & \cellcolor[rgb]{0.2563,0.5700,0.7752}{\color[rgb]{1,1,1}\textbf{4.115}} & \cellcolor[rgb]{0.3566,0.6393,0.8146}{\color[rgb]{0,0,0}\textbf{3.611}} & \cellcolor[rgb]{0.2279,0.5416,0.7605}{\color[rgb]{1,1,1}\textbf{4.306}} & \cellcolor[rgb]{0.2400,0.5538,0.7668}{\color[rgb]{1,1,1}\textbf{4.233}}\\
\hline
\end{tabular}
}
\end{table}

\begin{figure}[!htb]
\resizebox{\columnwidth}{!}{
\begin{tabular}{cc}
\includegraphics[width=0.48\columnwidth]{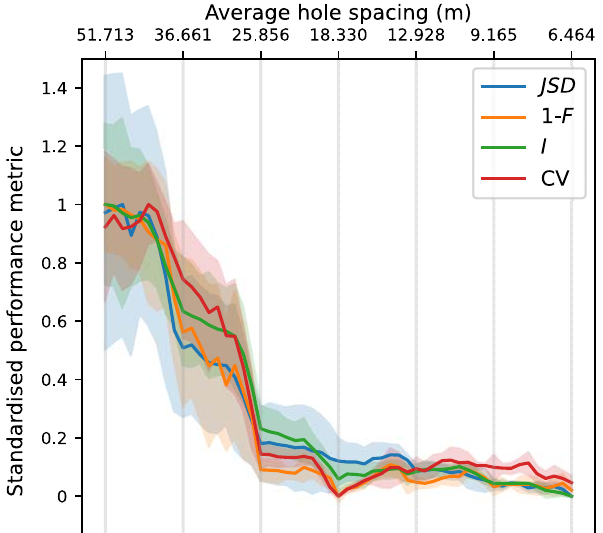} &
\includegraphics[width=0.48\columnwidth]{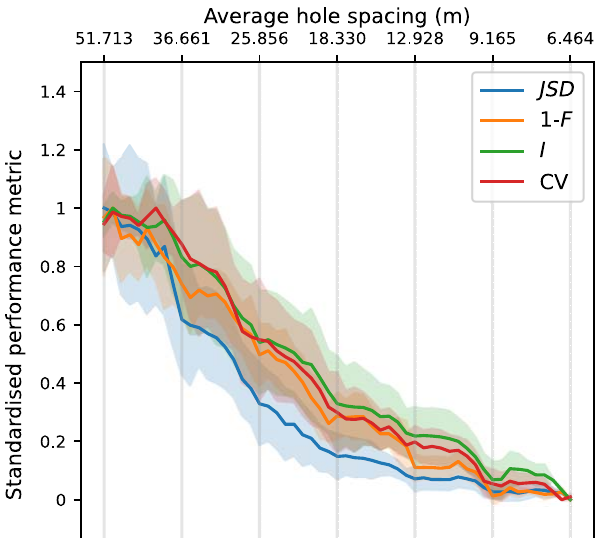}\\
\end{tabular}
}
\caption{Performance curves for (left) waste and (right) mineralized domains}\label{fig:all-performance-curves}
\end{figure}

\begin{figure}[!htb]
\centering
\includegraphics[width=0.6\columnwidth]{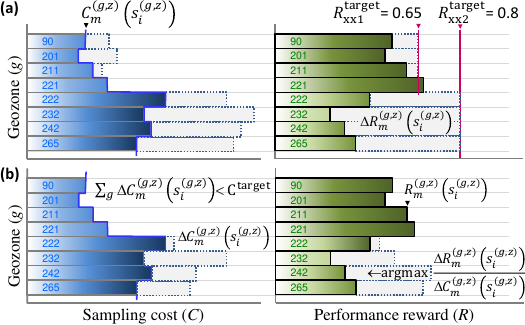}
\caption{Performance curve application scenarios}\label{fig:applic-performance-curves}
\end{figure}

In order to assess the speed of improvement, the normalized metrics are standardized according to (\ref{eq:standardized-metric}) to enable the rate of change w.r.t. $s$ to be directly compared between metrics.
\begin{align}
\check{\xi}_{m,s}^{(g,z_\text{RL})}=\frac{\xi_{m,s}^{(g,z_\text{RL})}-\min_s \xi_{m,s}^{(g,z_\text{RL})}}{\max_s \xi_{m,s}^{(g,z_\text{RL})} - \min_s \xi_{m,s}^{(g,z_\text{RL})}}\in[0,1]
\label{eq:standardized-metric}
\end{align}
Fig.~\ref{fig:individual-performance-curves} shows the \textbf{performance curves} obtained for individual metrics. It emphasizes the need to treat mineralized\,/\,hydrated, and waste domains separately due to differences in their convergence behavior, especially in relation to fidelity and uncertainty. In Fig.~\ref{fig:all-performance-curves}, the distortion, fidelity and uncertainty performance curves (JSD, F, I and CV) are shown together in a single plot for waste and mineralized domains. An important observation is that CV (the red curve) can serve as a proxy measure for estimating the rate $\Delta\check{\xi}_{m}^{(g)}(s)/\Delta s$ of the other three. Crucially, CV can be computed without knowledge of the {ground truth}. Thus, a fitted model, $\tilde{y}^{(g)}_m(s\!\mid\!\check{\xi}^{(g)}_{m,s})$, can provide a useful bound for guiding incremental sampling decisions. Assuming a region has area $A^{(g,z_\text{RL})}$ and sampling density $\lambda(s_i)$, the \textbf{incremental cost and reward} for moving from $s_i$ to a denser state $s_{i+1}$ would be
\begin{equation}
\Delta C(s_i\!\rightarrow\! s_{i+1})\propto A^{(g,z_\text{RL})}\cdot\left[(1+f_{i+1})\lambda(d_{i+1})-(1+f_i)\lambda(d_i)\right]\label{eq:deltaC}
\end{equation}
\begin{equation}
\Delta R(s_i\!\rightarrow\! s_{i+1})\propto w^{(g,z_\text{RL})}_m\cdot G^{(g,z_\text{RL})}_m\cdot\left[\tilde{y}^{(g)}_m(s_i) - \tilde{y}^{(g)}_m(s_{i+1})\right]\label{eq:deltaR}
\end{equation}
where $w^{(g,z_\text{RL})}_m$ and $G^{(g,z_\text{RL})}_m$ denote its worth and maximum gain based on the normalization factor in (\ref{eq:normalized-metric}) and Table~\ref{tab:potential-gains}. Fig.~\ref{fig:applic-performance-curves}(a) illustrates a scenario that focuses on achieving a fixed level of quality. This is articulated by two performance targets $R^\text{target}_{xx1}$ and $R^\text{target}_{xx2}$ which are set at 0.65 and 0.8, respectively, for waste and hydrated/mineralized domains. If the CV metric is chosen, this would mean setting the goal of achieving 65\% and 80\% of the total possible uncertainty reduction based on simulation or previous experience [e.g. from an adjacent pit/patch or the bench above]. The action of drilling additional holes is expected to yield a reward of $\Delta R^{(g,z)}_m(s_i\rightarrow s_{i+1})$ for each geozone. Since the scaling in (\ref{eq:normalized-metric})--(\ref{eq:standardized-metric}) are reversible, decisions can be related directly to geo-confidence goals (e.g. aiming for $<2$\% error) by mapping $\tilde{y}^{(g)}_m(s\!\mid\!\check{\xi}^{(g)}_{m,s})$ back to raw CV values, $\zeta^{(g)}_{m,s}$. Scenario (b) seeks to balance performance and determine the number of additional holes to drill in relevant geozones subject to a cost constraint. Through reinforcement learning \cite{levinson2024reinforcement}, an agent may find that geozone 242 returns the highest reward per unit cost.

\subsection{Experiment 2}\label{sec:results-exp2}
\begin{figure}[!htb]
\centering
\resizebox{0.6\columnwidth}{!}{
\setlength{\tabcolsep}{1pt}
\begin{tabular}{cc}
\multicolumn{2}{c}{\scriptsize Region A}\\
\includegraphics[width=0.25\textwidth]{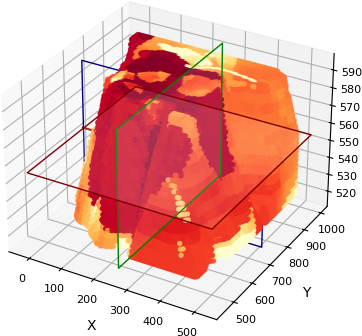} &
\includegraphics[width=0.25\textwidth]{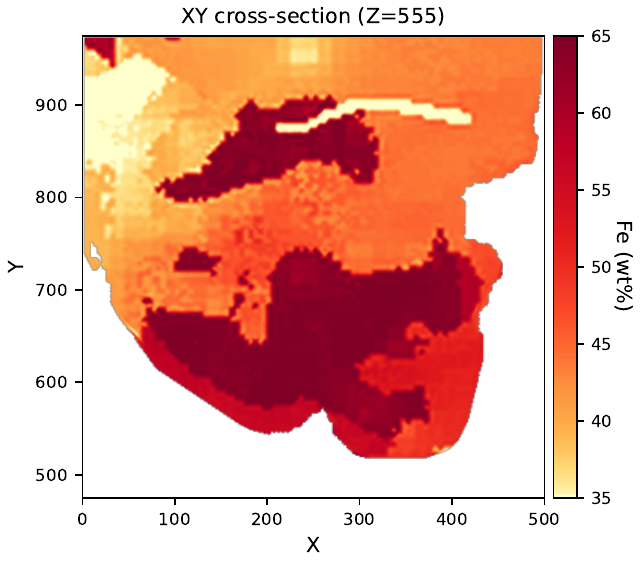}\\
\includegraphics[width=0.25\textwidth]{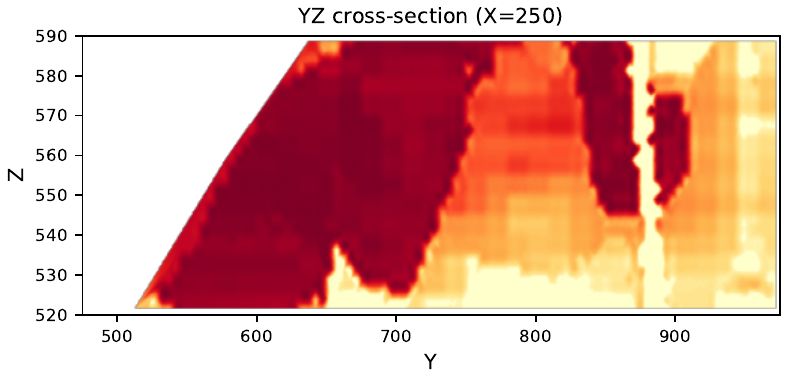} &
\includegraphics[width=0.25\textwidth]{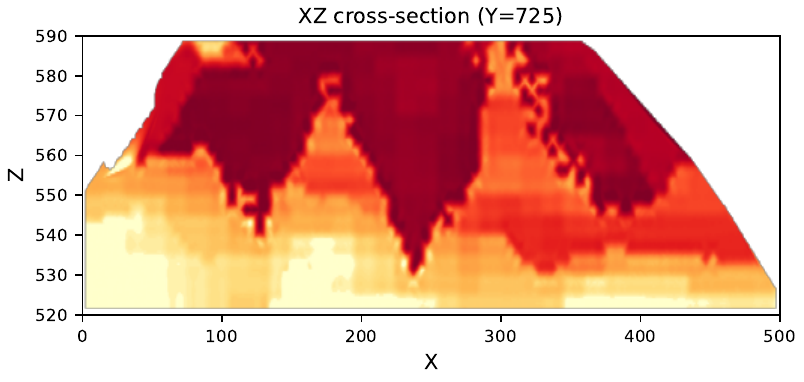}\\
\multicolumn{2}{c}{\scriptsize Region C}\\
\includegraphics[width=0.25\textwidth]{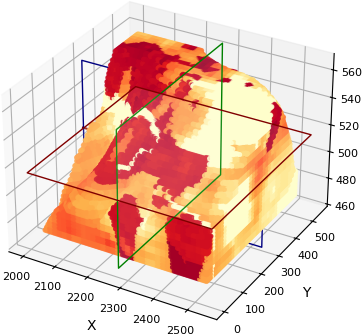} &
\includegraphics[width=0.25\textwidth]{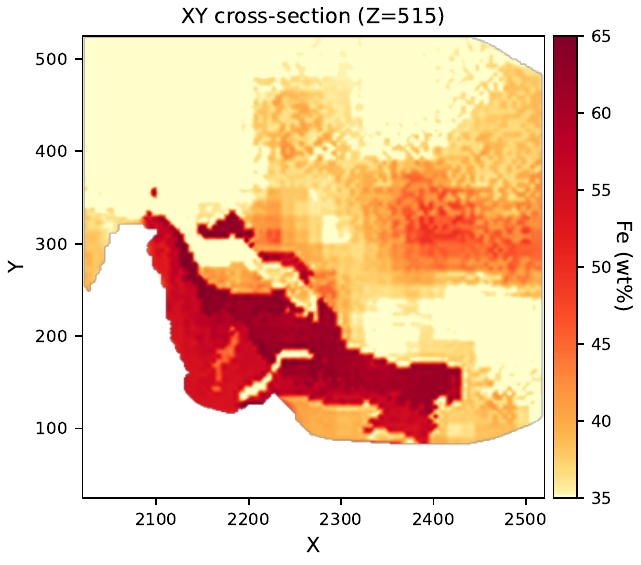}\\
\includegraphics[width=0.25\textwidth]{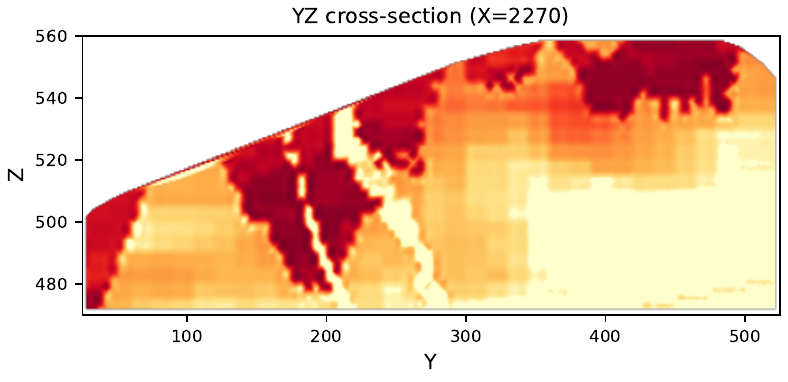} &
\includegraphics[width=0.25\textwidth]{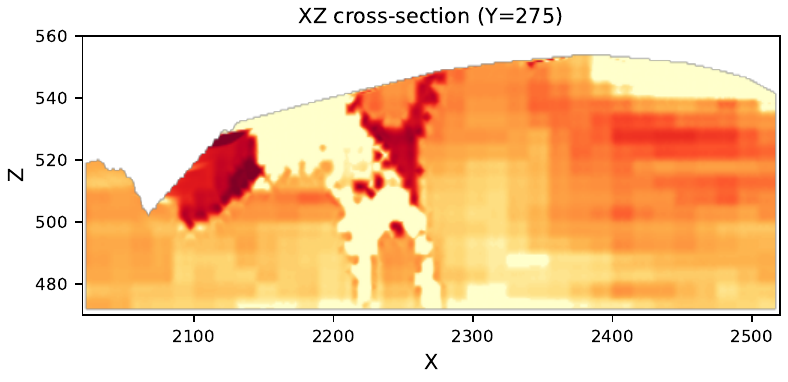}
\end{tabular}
}
\caption{Three-dimensional and cross-sectional views of two sub-volumes targeted in experiment 2. Both regions have high geological complexity}\label{fig:orebody-structure-complex-regions}
\end{figure}

Turning our attention to adaptive sampling in geologically complex sub-volumes, Fig.~\ref{fig:orebody-structure-complex-regions} shows the orebody composition (Fe concentration) in two targeted regions. In general, the red and orange-yellow pixels may be interpreted as mineralized and unmineralized zones, respectively, punctuated by igneous intrusion (thin veins). In the ensuing discussion, we focus first on one specific test case to demonstrate the potential of the adaptive sampling strategies. Subsequently, we examine general performance across four test regions.

\begin{figure*}[!htb]
\centering
\resizebox{\textwidth}{!}{
\setlength{\tabcolsep}{1pt}
\begin{tabular}{cccc}
\includegraphics[width=0.33\textwidth]{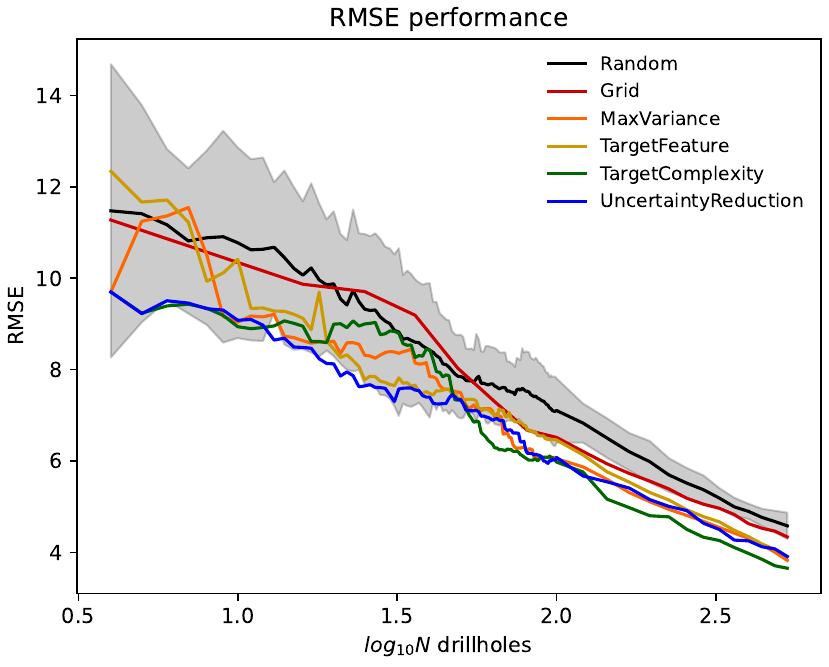} &
\includegraphics[width=0.33\textwidth]{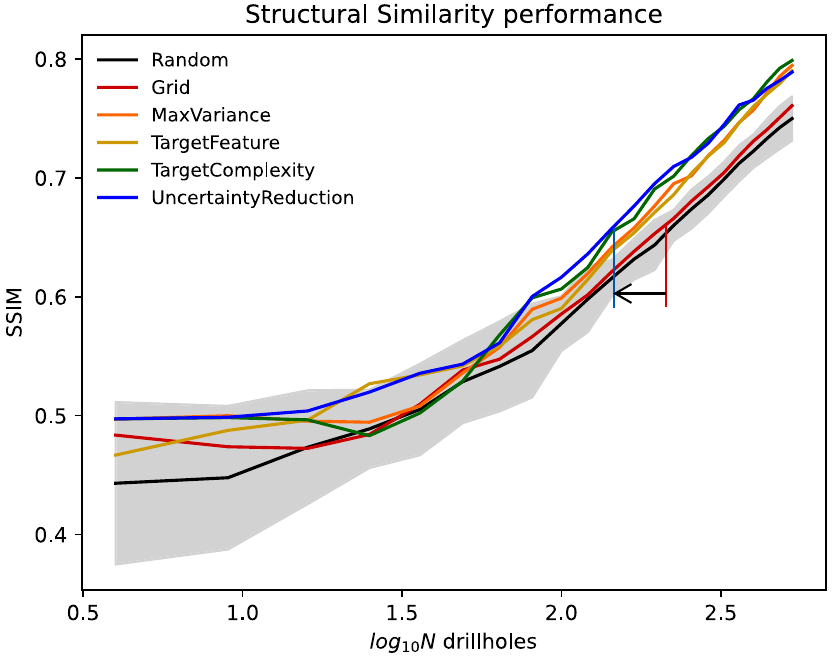} &
\includegraphics[width=0.33\textwidth]{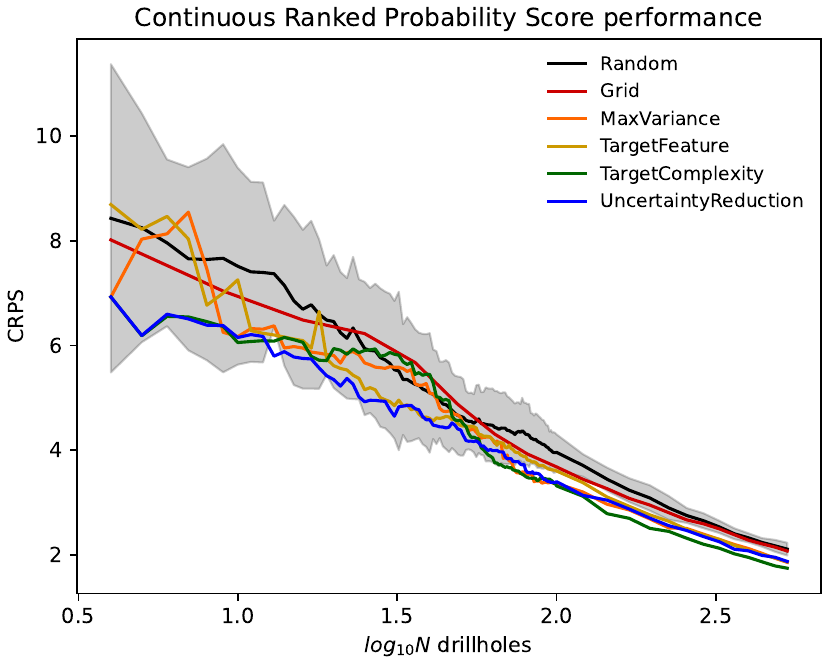}\\
\end{tabular}
}
\caption{Performance for different sampling strategies in region A and $z\in[570,590)$m. (Left) RMSE, (center) SSIM, (right) CRPS}\label{fig:adaptive-sampling-performance-curves-region0-RL570-590}
\end{figure*}

Fig.~\ref{fig:adaptive-sampling-performance-curves-region0-RL570-590} shows the performance of various strategies in region A at an elevation of 570-590m. The performance curves---for Random, Grid, MaxVariance, TargetFeature, TargetComplexity and UncertaintyReduction, abbreviated RN, GD, MV, TF, TC and UR---are plotted as functions of drill-holes in $\log_{10}n$. Stepping through the panels from left to right, performance is reflected through the RMSE, structural similarity (SSIM) and continuous ranked probability scores (CRPS). RMSE is arguably not the best measure as it treats prediction errors as uncorrelated. Nevertheless, we find that its trends often follow closely to that of CRPS which takes into account the estimated uncertainty. SSIM is analogous to the spatial fidelity measure used in experiment 1. Being a quality metric, it reveals structural similarity as its name suggests.

\textbf{Interpretation:} The black curve and shaded envelop illustrate the highly variable nature of the results attributed to random sampling (RN). On average, it ranks as the worst no matter which measure is used. The red curve (GD) shows that regular grid sampling is an improvement over RN. The caveat is that the chosen locations are inherently non-embedded, since two training sets of different sizes are chosen independently $(S_{n_1}\not\subset S_{n_2}$ for $n_1<n_2$).\footnote{Under GD, once we commit to a drill pattern spacing, it is generally not possible to move precisely to a higher density, increasing the number of drill-holes arbitrarily from $n_1$ to $n_2$ by drilling $(n_2-n_1)$ extra holes.} Thus, GD is merely a theoretical construct utilized as a convenient benchmark. The adaptive sampling strategies, starting with the orange curve (MV), consistently outperform GD. When $\log_{10}n\ge 1$, the RMSE distortion for MV, TF, TC and UR remained below GD. In general, UR performs better than MV. However, it is less effective than TC which seems to be the most robust. In fact, adaptive sampling maintains a sizable advantage over GD as $n$ grows for this specific test region, and this is particularly pronounced according to the SSIM fidelity measure.
\begin{itemize}
\item \textbf{Adaptive sampling gain:} Concretely, the TC vs GD quality gap measures up to 1.5 dB or 0.15 base-10 logarithmic units (see arrow). If this finding were to hold more broadly, it would equate to a 30\% reduction in drill-hole count, particularly in geologically complex areas.\,$\star$
\end{itemize}

\begin{figure}[!htb]
\centering
\resizebox{0.7\columnwidth}{!}{
\setlength{\tabcolsep}{5pt}
\begin{tabular}{ccc}
\textbf{Groundtruth} & \textbf{Regular Grid (GD)} & \textbf{MaxVariance (MV)}\\
& {SSIM=1.022, CRPS=0.966} & {SSIM=1.055, CRPS=0.911}\\
\includegraphics[width=0.33\columnwidth]{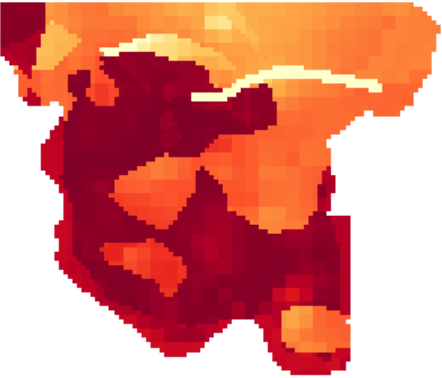} &
\includegraphics[width=0.33\columnwidth]{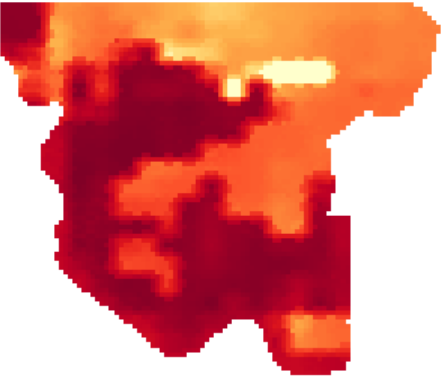} &
\includegraphics[width=0.33\columnwidth]{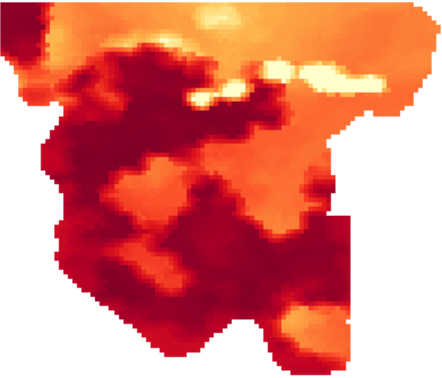}\\
\textbf{TargetFeature (TF)} & \textbf{TargetContrast (TC)} & \textbf{UncertaintyReduction (UR)}\\
{SSIM=1.068, CRPS=0.820} & {SSIM=1.060, CRPS=0.865} & {SSIM=1.074, CRPS=0.847}\\
\includegraphics[width=0.33\columnwidth]{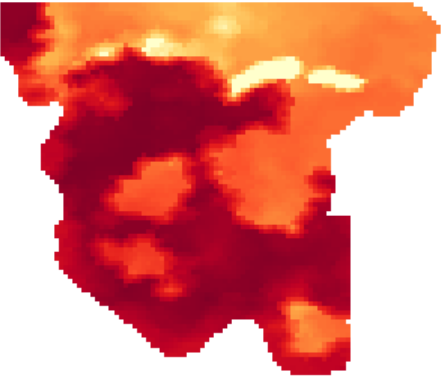} &
\includegraphics[width=0.33\columnwidth]{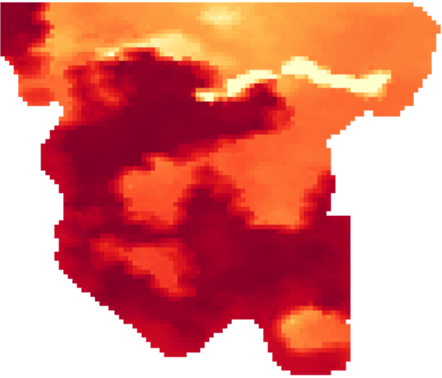} &
\includegraphics[width=0.33\columnwidth]{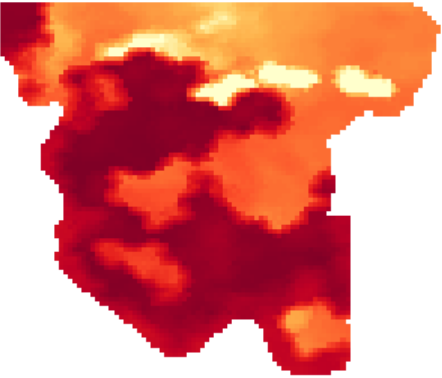}\\
\end{tabular} 
}
\caption{Sampling strategies comparison: region A, 570-590m ($n=400$)}\label{fig:visual-comparison-sampling-strategies}
\end{figure}

\newpage
A qualitative comparison is provided in Fig.~\ref{fig:visual-comparison-sampling-strategies}, showing that adaptive sampling strategies are, at least in this instance, more capable of preserving spatial structures in a complex geological setting than random and grid sampling.

\begin{table}[!htb]
\caption{Aggregate performance of sampling strategies in geologically complex regions, $\chi^{\Omega}_{m,\pi}$}
\centering
\resizebox{0.5\columnwidth}{!}{
\renewcommand{\arraystretch}{1.0}
\begin{tabular}{|l|cccccc|}
\hline
Strategy $\pi$ & RN & GD & MV & TF & TC & UR\\\hline\hline
Measure $m$ & \multicolumn{6}{c|}{Region A}\\\hline
\hspace{3mm}RMSE & 1 & 0.924 & 0.902 & 0.903 & \textbf{0.883} & 0.893\\ 
\hspace{3mm}CRPS & 1 & 0.917 & 0.906 & 0.900 & \textbf{0.888} & 0.891\\ 
\hspace{3mm}SSIM & 1 & 1.042 & 1.046 & 1.052 & 1.051 & \textbf{1.055}\\\hline 
Measure & \multicolumn{6}{c|}{Region B}\\\hline
\hspace{3mm}RMSE & 1 & 0.915 & 0.966 & 0.952 & \textbf{0.852} & 0.949\\
\hspace{3mm}CRPS & 1 & 0.911 & 0.982 & 0.954 & \textbf{0.874} & 0.948\\
\hspace{3mm}SSIM &  1.008 & \textbf{1.033} & 1 & 1 & 1.022 & 1.018\\\hline
Measure & \multicolumn{6}{c|}{Region C}\\\hline
\hspace{3mm}RMSE & 1 & 0.921 & 0.918 & 0.917 & \textbf{0.902} & 0.916\\
\hspace{3mm}CRPS & 1 & \textbf{0.906} & 0.926 & 0.929 & \textbf{0.908} & 0.922\\
\hspace{3mm}SSIM & 1 & \textbf{1.056} & 1.052 & 1.047 & 1.048  & 1.047\\\hline
Measure & \multicolumn{6}{c|}{Region D}\\\hline
\hspace{3mm}RMSE & 0.955 & 1 & 0.863 & 0.853 & \textbf{0.838} & 0.848\\
\hspace{3mm}CRPS & 0.905 & 1 & 0.861 & 0.862 & 0.829 & \textbf{0.815}\\
\hspace{3mm}SSIM & 1.003 & 1 & 1.025 & 1.034 & \textbf{1.039} & 1.032\\\hline
\end{tabular}
\label{tab:rmse-crps-ssim-scores}
}
\end{table}
To assess overall performance, the measures are computed over the entire z interval (ranging from 80 to 120m) for each test region. The area under each performance curve, $\zeta^{\Omega}_{m,\pi}(n)$, is then integrated and normalized to obtain a score $\chi^{\Omega}_{m,\pi}$ for each measure, $m\in\{\text{RMSE, CRPS, SSIM}\}$, strategy $\pi$ and region $\Omega$ as shown in (\ref{eq:scenario2-scores}). These are reported in Table~\ref{tab:rmse-crps-ssim-scores}.
\begin{equation}
\chi^{\Omega}_{m,\pi}\propto\int_{n\ge n_{\min}}\zeta^{\Omega}_{m,\pi}(n) dn,\quad\text{ with }n_{\min}=9\label{eq:scenario2-scores}
\end{equation}

\textbf{Discussion:} The main contribution of experiment 2 is demonstrating the potential of employing adaptive sampling (AS) in geologically complex areas. The scope and design of this study are unique, as it examines four heterogeneous regions (up to 500$\times$500$\times$120m$^3$ in size), four adaptive sampling strategies (MV, TF, TC and UR) using three performance measures (RMSE, CRPS and SSIM). It shows more efficient information gathering is not only \textit{possible}, but a combination of GP regression with the TargetComplexity strategy, for instance, provides a viable way of achieving this (see learning and inference steps in Sec.~\ref{sec:conceptual-framework} and algorithm prescription in Sec.~\ref{sec:methods}). Though it must be said, the performance of these adaptive sampling strategies can change depending on the starting configuration. In this paper, a $2\!\times\!2$ drill pattern [with padding and one hole in each corner] is used for initialization before adaptive sampling begins. Preliminary result suggests TF performance can improve further using a denser configuration. This is likely due to a combination of factors, such as volatility in the estimated kernel parameters, and inaccuracy in the signal gradient estimates when $n_0$ (the initial number of drill-hole samples) is small. AS sensitivity to the starting configuration may be further investigated in future work.

\newpage
\textbf{Significance:} The key observation (see $\star$ above) is that adaptive sampling can reduce drilling requirement in complex regions, for a given level of prediction error or structural similarity, by up to 30\% in some instances relative to regular grid sampling. While this finding is preliminary, and the gain might not be sustained at scale across deposits, it is still a noteworthy result. According to industry analysts \cite{maptek2020mt}, a 5\% reduction in over-drilling can lead to savings of \$0.5M in just a few months. Thus, the finding is significant (with benefits estimated in the tens of millions) even when modeled on a 10\% efficiency gain.\footnote{For 2026, Pilbara iron ore exports are forecasted at 890-900 million tonnes (Mt). According to BHP and Rio Tinto's cost guidance, the total unit operating cost (C$_1$) is estimated at US\$18-23 per tonne. Drilling typically accounts for 5-8\% of the total mining cost, with assaying adding a further 25\% to this cost. If drill-hole saving (DHS) is assumed to be 10\% and realized for 20\% of all drilling, a conservative estimate of the benefits of adaptive sampling would be valued at US\$\,20-40M per annum (based on 890[Mt]$\times$18[US\$/t]$\times$(0.05$\times$1.25)[drilling+assays]$\times$0.1[DHS]$\times$0.2). Possibly higher if DHS and realization ratios are higher.}

\section{Concluding Remarks}
This paper presented a conceptual framework for evaluating the value of drill-hole information using Gaussian Process and multiple statistics. Resampling basically repurposed a matured model [the high-resolution grade estimation blocks] for data mining. In this contribution, the experiments catered for situations where geological domains are differentiated or mixed. In scenario 1, performance curves were obtained to inform in-fill drilling and spacing consideration consistent with current practice. Using the coefficient of variation as a proxy measure, it is possible to estimate the incremental cost and reward without knowing the {ground truth} using insights obtained from a similar geozone or adjacent bench \cite{leung2026arxiv-supp}. Scenario 2 provided clear evidence of situations where regular grid sampling is suboptimal and how performance can be improved. TargetComplexity was determined as an adaptive sampling strategy capable of reducing drilling requirement in geologically complex regions, and being applied in the future as mining robotics, AI and ML technologies continue to evolve. Future work will provide more extensive analysis, delve deeper into the conceptual foundations and explore connections with active learning and other fields.

\bibliographystyle{unsrt}
\bibliography{references.bib}

\end{document}